# Primary accelerometer calibration with two-axis automatic positioning stage

*Wataru Kokuyama* [a,*], *Tomofumi Shimoda* [a] *and Hideaki Nozato* [a]

[a] National Metrology Institute of Japan (NMIJ), National Institute of Advanced Industrial Science and Technology (AIST), Tsukuba, 305-8563 Japan.
* Corresponding author, wataru.kokuyama@aist.go.jp

*Abstract* − In this study, we developed an automated, multipoint primary accelerometer calibration system using a two-axis positioning stage and a heterodyne laser interferometer. The proposed system offers low-cost, convenient, and automated multipoint accelerometer calibration, enabling less calibration lead time. The positioning stage also offers better positioning repeatability of 1 μm, which is impossible through manual alignments. We measured the surface deformation of a laser reflection adaptor for a single-ended accelerometer by measuring more than 450 measurement positions. Visualizing the deformation of laser reflection surfaces facilitates understanding the effects of deformation or nonrectilinear motion, which are among the most significant uncertainty components in high-frequency accelerometer calibrations.

*Keywords:* Accelerometer calibration, Vibration calibration, Laser interferometer, Positioning stage.

## 1. INTRODUCTION

The primary calibration of accelerometers, standardized in ISO 16063-11:1999 [1], is conducted in many national metrology institutes to provide reliable vibration measurements for various industries, such as automobiles. The National Metrology Institute of Japan (NMIJ) also provides national vibration standards up to 10 kHz. Also, it participates in several international comparisons to maintain international measurement equivalence. Recently, vibration standards exceeding 10 kHz are needed, especially after the recent boom in the electric vehicle industry, in which high-frequency vibrations are generated by electric power trains.

In the range of such high frequencies, more than 3–4 positions should be used for the displacement measurements with laser interferometers, as accelerometers have undesired motions, such as nonrectilinear motions. In the international key comparison CCAUV-V-K5 [2], the protocol requires that "the measurements should be performed for at least three different laser positions, which are symmetrically distributed over the respective measurement surface." Measurements with several laser positions are time-consuming and complicated, especially with manually aligned lasers. Additionally, manual alignments may have inaccurate positioning reproducibility.

Some techniques are independent of manual alignment. For example, a galvanometer scanner realizes laser scanning with a single-point heterodyne laser interferometer [3]. Commercial scanning laser Doppler vibrometers (LDVs) also offer multiple-point measurements. However, scanning LDVs are generally expensive and inflexible in customizing signal processing and mechanical configurations. Another option entails multipath interferometers, which can simultaneously measure multiple points to reduce the deformation effect [4]. However, such interferometers require complicated alignment operations to spot laser beams to target the required measurement positions.

Herein, to realize a low-cost, convenient, and automated multipoint accelerometer calibration system, we used a single-point laser interferometer and a two-axis positioning stage to automatically point the laser beams at any position. Utilizing the ability to measure many positions quickly, we measured and visualized the surface deformation of a laser reflection adaptor with more than 450 positions. The results showed that a hexagonal-shaped adapter is more suitable for using transfer artifacts than a cylindrical adapter with cutting. Herein, we introduced the proposed system and explained the measurement results. This article is an extended and revised version of a proceedings paper [5] presented at the IMEKO 2021 online conference.

## 2. CALIBRATION SYSTEM

Fig. 1 presents the block diagram of the calibration system. This system is based on the primary vibration calibration method with a sine approximation, defined in method 3 of ISO16063-11:1999 [1]. The accelerometer under measurement (device under test (DUT)) is attached to a metrology-grade vibration exciter (Endevco, model 2911 or Spektra, model SE-09) with beryllium- or ceramics-made armature. The excitation current is injected by a power amplifier (Spektra, BAA500), and the excitation signal is generated by a signal generator (NF cooperation, WF1948), which is controlled using a personal computer (PC).

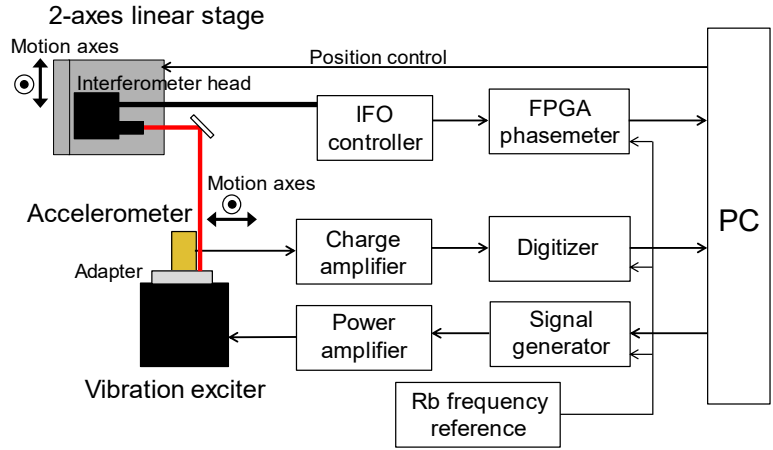

Fig. 1. Block diagram of the system. PC, FPGA and IFO represent personal computer, field-programmable gate arrays, and interferometer optics, respectively.

We used a commercial heterodyne laser interferometer (Ono Sokki, LV-9002) with a He–Ne laser (633-nm, < 1 mW) and a variable focus lens as the displacement measuring instrument. The frequency of the He–Ne laser is stabilized within a relative uncertainty of $\pm 2 \times 10^{-7}$ using two-mode frequency stabilization. The absolute wavelength in a vacuum is specified as 632.8194 nm. The focal length of the output beam can be varied from 10 cm to infinity. However, it was fixed during the experiment because the optical distance does not relatively change, even in the case of low-frequency (< 10 Hz) vibrations; the vibration amplitude is less than millimeters. In our setup, the optical length is ~70 cm, and the laser beam is focused to a diameter of 70 μm. The direction of the output beam from the interferometer is changed from horizontal to vertical with a wide-diameter mirror. The interferometer head is mounted on a two-axis linear stage (Fig. 2(b)), which is made of vertical (Kohzu, ZA16A-X1) and horizontal (Kohzu, XA16A-X1) stages. The stages are based on a cross roller guide, with a motion range of ±25 mm, which sufficiently covers the area of the accelerometer or laser reflection adapter. According to the datasheet, the positioning resolution is 1 μm, and the positioning repeatability is < ±0.5 μm. The accumulated lead error, defined as the maximum position error among the full motion range, is specified as < 10 μm in 50 mm. The vertical stage can retain up to 10-kg weight, sufficient for holding the laser interferometer and its jig, weighing approximately 3 kg in total.

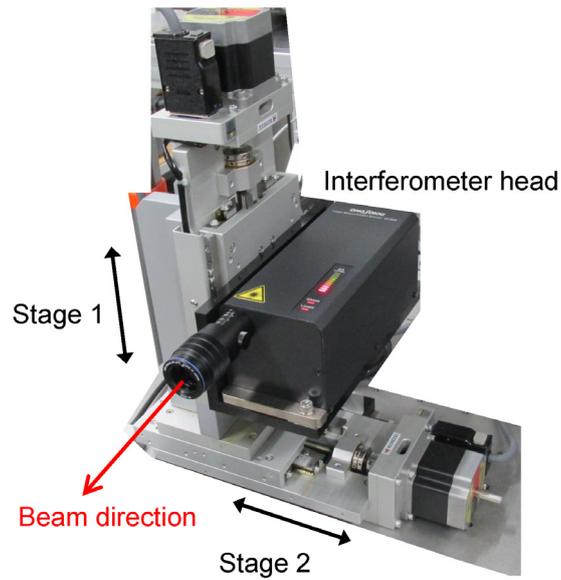

Fig. 2. The head of heterodyne laser interferometer on a two-axis linear stage.

The interferometer controller outputs the 80-MHz heterodyne beat signal, as well as the reference signal which drives the acousto-optic modulator (AOM) inside the interferometer head. To obtain the displacement signal, we used a custom-made field-programmable gate array (FPGA) phase meter [6], based on zero-crossing counting so that the phase difference of the fast-modulated heterodyne signals can be precisely measured. The algorithm was invented by our group, and we confirmed that the phase-measurement error is $\sim 1 \times 10^{-4}$ rad [6], which is sufficient for this measurement. The data is sent to the PC via a USB 3.0 link, and the sampling rate of the phase measurement can be set to the desired rate between 10 Hz and 250 MHz. Note that the Heydemann correction [7], used to reduce the cyclic errors of laser interferometers, was not applied. Instead, a technique with a background motion [8] was applied to eliminate the cyclic error.

When a single-ended accelerometer is the DUT, the laser reflection adapter is used. The adapter is made of stainless steel (SUS316L), and its top surface is polished to reflect the laser beams. According to JIS B0601-2001 [9], the mean surface roughness $(R_\mathrm{a}) \simeq 0.7$ nm and ten-point mean surface roughness $(R_\mathrm{zJIS}) \simeq 5$ nm. Herein, two shapes were prepared for comparison: a hexagonal and a cylindrical adapter with cutting. The latter has exactly the same shape as that used in CCAUV.V-K5. The two contact surfaces, between the accelerometer and the adapter, and between the adapter and the armature, are lubricated with ultrahigh vacuum grease (M&I materials, Apiezon L). The mounting torque is fixed to 2.0 Nm using a calibrated torque wrench.

A charge amplifier (B&K, 2692-A-0S4) converts the charge output from the accelerometer to voltage. Between the accelerometer and charge amplifier, a 1.2-m super low-noise cable (B&K, AO 0038-D-012) is used to connect them. The voltage output from the charge amplifier is measured using a digitizer (National Instruments, NI PXI-5922), with a 24-bit resolution and 15-MSps (sample per seconds) maximum sampling rate. The gain of the input channel of the digitizer is calibrated in our laboratory using a standard voltage reference (Yokogawa, GS210), which is calibrated with uncertainty of $2 \times 10^{-5}$ ($k = 2$). The digitizer calibration is conducted using DC (0 Hz), and the measurement has a deviation of $< \pm 0.05\%$.

The two acquired datasets for the interferometer and accelerometer voltage signals are applied with sinusoidal curve fitting (or sine approximation); the amplitude and phase of each sinusoid are obtained. The fitted frequency is the vibration frequency. According to the sine-approximation method in ISO16063-11, the sensitivity (magnitude) and phase shift for the specified frequency can be obtained by comparing both fitting results. In this curve fitting, a technique [10] comprising the two following processes was applied. First, the interferometer signal is a second-order differential of raw displacement signal to match its dimension to acceleration to cancel out the common background noise. Second, the Hanning window is applied before the fittings to avoid spectral leakage of noises from far separated frequencies from the vibration frequency. The details of this fitting technique are described and analyzed in another study [10].

The FPGA phase meter, the digitizer, and the signal generator are locked to the same rubidium (Rb) frequency reference (Stanford Research Systems, FS725) to ensure traceability to the SI time on the order of $10^{-12}$. Furthermore, the FPGA phase meter and digitizer are triggered by the same signal so that the acquisition timing of the two instruments is matched.

## 3. EXPERIMENTAL RESULTS

### 3.1. Repeatability

One of the concerns of using the positioning stage is stability. Thus, we checked whether the stage deteriorates the repeatability of the measurements. The single-ended-type standard accelerometer (B&K, type 8305-001) was used as a DUT. In this measurement, we chose a hexagonal-shaped adapter. Fig. 3(a) shows the picture of the connected accelerometer.

Six different points in the laser reflection adapter were selected for measurement (Fig. 3(b)). In Fig. 3(b), the color plot shows the intensity distribution of the laser interferometer output, showing the reflected optical power from the targeted position. The two-dimensional map is obtained by monitoring the intensity of the laser interferometer signal while scanning the position of the laser interferometer. Red represents the maximum intensity, corresponding to the adapter's mirror surface. Here the accelerometer's top surface also shows maximum intensity; however, the area is not suitable for measurement and is not used. The six-measurement points, shown as black-filled circles in Fig. 3(b), are aligned with the accelerometer's apexes. The positions were point-symmetrical to the center of the accelerometer so that the effect of the nonrectilinear motion is minimized when averaged. For comparison, the top view of the accelerometer's photo is also shown in Fig. 3(c).

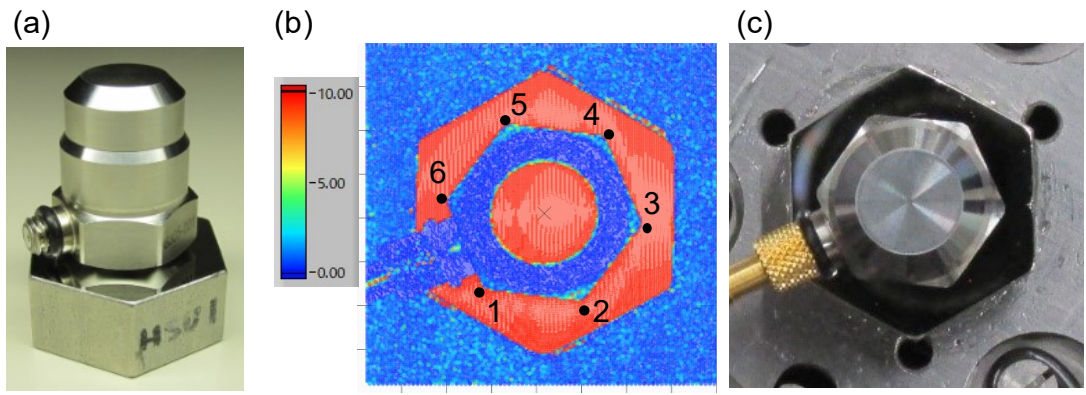

Fig. 3. (a) Accelerometer with the laser reflection adapter, (b) six-measurement point sets overlayed on an interferometer output intensity map, and (c) top view of the device under test.

Calibration results were obtained for the following conditions, and the standard deviations were also calculated: (i) single point (N = 6), (ii) six different points, and (iii) average of the six points (N = 6). The vibration frequencies at or under 10 kHz are a third-octave standard sequence defined by ISO266:1997 [11]. Above 10 kHz, the frequency interval is 500 Hz. The vibration amplitude for calibration is a standard sequence used in NMIJ (e.g., $(0.5, 5, 10, \text{and } 100)$ m/s$^2$ at $(5, 20, 160, \text{and over } 10000)$ Hz, respectively), which is the same amplitude used in CCAUV.V-K5. Note that any desired frequencies and vibration amplitudes can be applied in this system.

The settings of the charge amplifier are the same as those used in the conventional procedures of calibration services: output unit = 10 mV/(ms$^{-2}$), high-pass filter = 0.1 Hz, low-pass filter = 100 kHz, and transducer sensitivity = 0.1 pC/(ms$^{-2}$). The ambient temperature of the laboratory is maintained at 21℃ ± 1℃ for 24 h. The humidity is also controlled at around 50%.

The magnitude results are shown in Fig. 4(a). The single-point result originates from the intrinsic repeatability of the measurement system, which is mainly from the random noise. From 10 Hz to 5 kHz, the single-point results are below 0.01%, showing sufficient low dispersion. The exception is 160 Hz, around which the vibration exciter has strong resonance. Below 10 Hz, the single-point results gradually increase because of the low signal-to-noise ratio (SNR) due to the limited acceleration level of the vibration exciter. Over 5 kHz, the single-point result also increases because of the increased noise level. The noise level in the measurements is analyzed in Subsection 3.2.

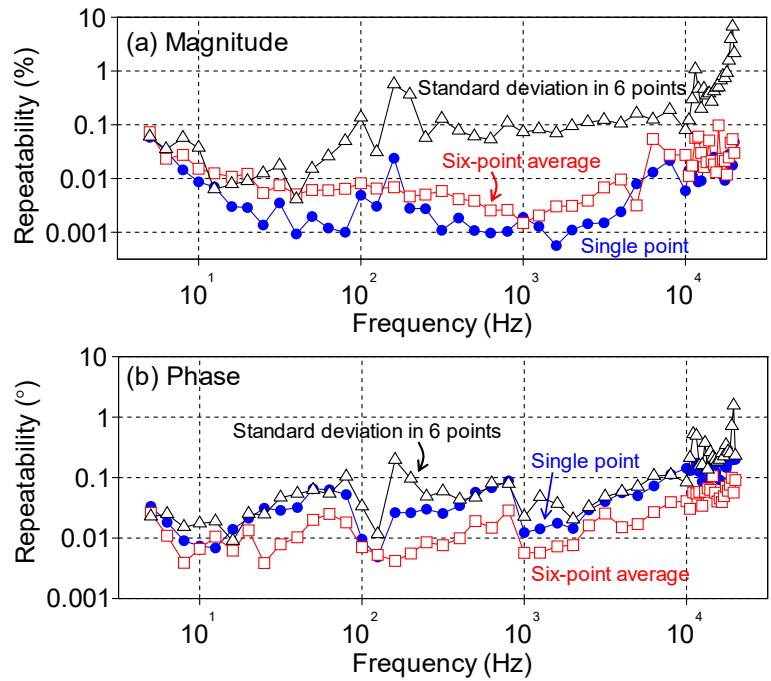

Fig. 4. Repeatability measurements for (a) magnitude and (b) phase.

In almost all frequency ranges, the six-point result, which is shown as black triangles in Fig. 4(a), is larger than the single-point results. This trend indicates undesired motions of the vibration exciter or the adapter, such as resonance, rocking motion, or deformation [12]. In the frequency range of approximately 100 Hz to 10 kHz, a rocking motion was found to cause a deviation of roughly 0.1%. Above 10 kHz, the deviation increases and reaches several percent because many resonances and deformation of the adapter arise. Note that the adapter's symmetric deformations may be suppressed and be underestimated in the six-point results because the six points are point-symmetric to the center. To determine the distribution, a scanning-like measurement is presented in Subsection 3.3.

The six-point average results are shown as red squares in Fig. 4(a) and (b), representing the suppression of the undesired motion by averaging. In the whole frequency range, the results are below 0.1%, which is suitable considering that the total uncertainty of accelerometer calibration is around 0.3%–0.4%. The remaining dispersion probably originates from the remnant of averaging, variation of undesired motions, and the accelerometer's stability. Notably, the averaging suppresses the dispersion by two orders of magnitude from the six-point standard deviation due to the precise positioning capability of this system. Additionally, we conclude that the positioning stage does not deteriorate the repeatability.

The dispersion of phase (Fig. 4(b)) shows simpler characteristics. This is because the rocking motion generally does not change the phase of the vibration, as it has no hysteresis in motion. The six-point results (black triangles) approximately match the single-point results (blue-filled circles), indicating that the phase has no position dependency. Exceptions exist around 160 Hz and over 10 kHz, caused by the strong resonance of the vibration exciter and the adapter's deformation.

Another origin of the phase deviation is the trigger mismatch (or skew) between the two signals. In this system, the sampling rate of data acquisition is increased at 100 Hz and 1 kHz, corresponding to a sudden decrease in the graph. This trend occurs because the trigger mismatch is inversely proportional to the sampling rate, and the phase deviation depends on the sampling rate. Note that this is simply from an inadequate design of the data acquisition software and will be fixed. By taking six-point average (red squares), the repeatability is < 0.1°, sufficient for the total combined uncertainty of calibration.

### *3.2. Noise level*

The origin of the dispersion of single-point measurement is noise. Thus, analysis of noise spectra provides useful information on the noise mechanism and performance of the calibration system. In Fig. 5, the noise spectra for the laser interferometer and accelerometer signals are plotted. The two vertical axes are aligned so that the vibration peaks are exactly overlapped. The displacement of the interferometer was converted from displacement to acceleration by taking the second derivative.

Fig. 5(a) shows the noise spectra for measurement at 5 Hz, at which the vibration amplitude is 0.5 m/s$^2$. At this frequency, the SNR of the voltage signal is low; the amplitude of the voltage output from the charge amplifier is only 5 mV. The sampling rate and measurement time were 100 kHz and 10 s, respectively, *i.e.*, 50 waves of the 5-Hz vibration were recorded. At this vibration frequency, the noise floor is higher in the accelerometer signal than that in the laser interferometer signal due to the low SNR. Thus, the measurement repeatability at this frequency originates from noise in the accelerometer signal.

From approximately 30 Hz to 2 kHz, the two spectra match, indicating that the signal is limited by common vibration noise applied to the accelerometer, such as external noise and noises in the power amplifier. In this case, the common noise does not affect the measurement dispersion ("single point" in Fig. 4) [10], owing to the second-order differentiation before the sinusoidal fitting. Thus, the dispersion is at its lowest level, around 0.001 %, in this frequency range.

Above ~2 kHz, the noise of the laser interferometer exceeds that of the accelerometer because of white displacement noise, which shows $f^2$ dependency in the acceleration dimension. Here, two mild peaks (around 30 Hz and 90 Hz), at which the noise of the laser interferometer exceeds that of the accelerometer, are observed. These peaks originate from resonances of the optical displacement measurement chain: mirrors, the laser interferometer, and the two-axis positioning stage. However, in these frequencies, the dispersions are undeteriorated compared to other frequencies, showing that this system's configuration does not have a deficiency. Note that if we use smaller vibration amplitude for calibration than the normal one ($10 \text{ m/s}^2$), the effect of the vibration noise in the optical displacement measurement chain may arise.

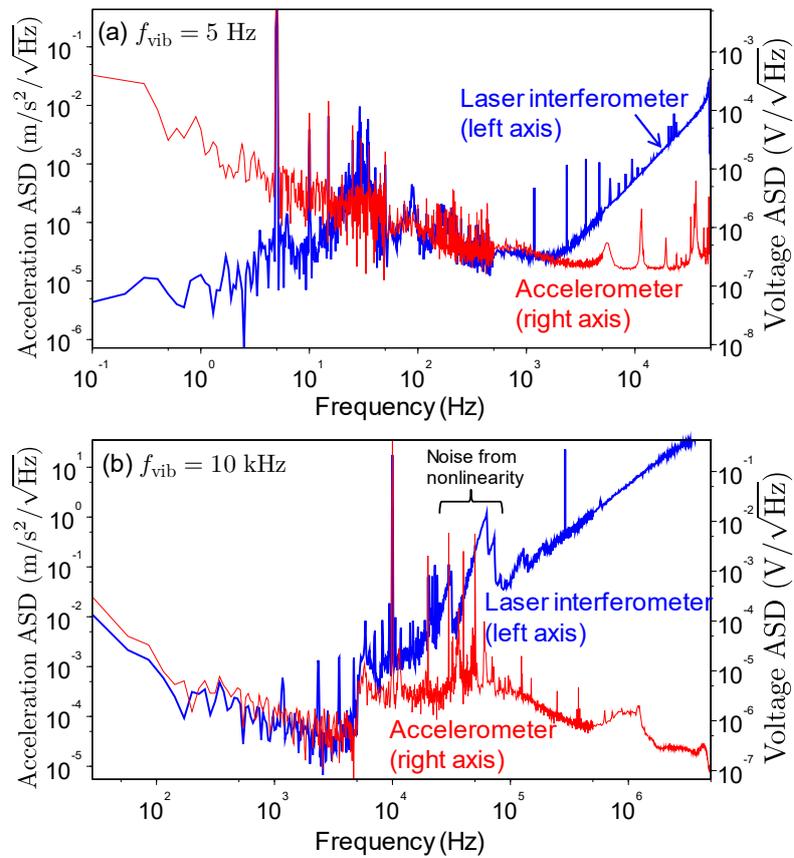

Fig. 5. Noise levels of the laser interferometer and accelerometer output. (a) $f_{\text{vib}}$ (vibration frequency) is 5 Hz, and (b) $f_{\text{vib}}$ is 10 kHz.

The noise spectra at the 10-kHz vibration frequency (Fig. 5(b)) show the characteristic in the higher-frequency range. Here, the background-motion technique [8] for eliminating the nonlinearity of the laser interferometer is applied; the vibration exciter's motion is a superposition of a 10-kHz sinusoidal signal for the main vibration and a 10-Hz triangular signal to impose background motion. Then, the center of triangular displacement, which shows an approximate linear motion, is selected for the data analysis. Note that the background motion does not affect the fitting of the sinusoidal vibration because the second-order derivation is applied before the sinusoidal fitting, eliminating low-frequency components, including drifts.

The technique causes a slight difference in the noise level in Fig. 5(b) from Fig. 5(a). First, the noise spectrum of the laser interferometer decreased in the range of 2 kHz–5 kHz. This trend occurs probably because the noises from the nonlinearity of the laser interferometer are converted to higher frequencies of 30 kHz–80 kHz. The nonlinearity makes several triangular-shaped peaks in the spectrum. Second, the noise spectrum of the accelerometer signal above 5 kHz increased. We did not specify the origin. Note that over 100 kHz, the accelerometer signal is suppressed by the low-pass filter in the charge amplifier; thus, such a higher-frequency region is meaningless.

### *3.3. Visualization of the surface deformation of laser reflection adapters*

Using the automatic positioning stage, several hundreds of positions can be measured in sequence to visualize the deformation of the laser reflection surface, which affects the calibration results significantly in the high-frequency range (>10 kHz). For example, we measured the deformation at 10 kHz for different laser reflection adapters (Fig. 6). The measurement positions are honeycomb-liked, equally spaced by 0.3 mm, and the measurement points are 466 and 468 for (a) and (b). The nominal vibration amplitude is 100 $\text{m/s}^2$.

The measurement is conducted in sequence to obtain the voltage sensitivity for each position. Then, the inverse of the voltage sensitivity is the displacement variation shown in Fig. 6, assuming that the accelerometer's sensitivity does not change during measurement because the measurement is conducted after sufficient warm-up. Such a procedure was used because the vibration amplitude fluctuates at about 1% as the vibration exciter is operated in an open-loop mode, not in a closed-loop control for the vibration amplitude. The color map plots of Fig. 6(a) and (b) show the deviation of the displacement amplitude for each laser reflection adapter. The origin (0% in the plot) is the average value of six-measurement positions for the hexagonal adapter.

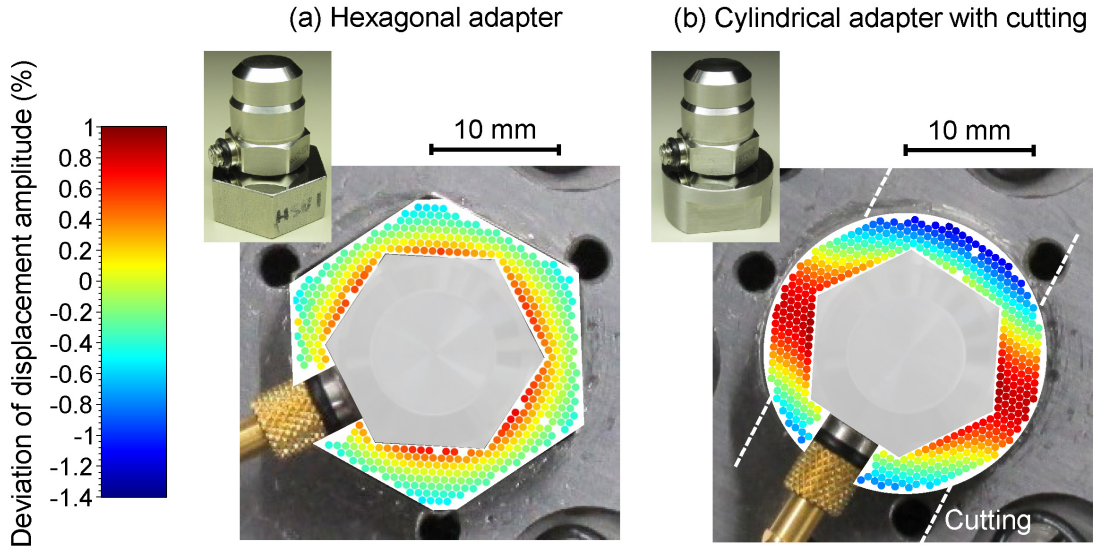

Fig. 6. Visualization of surface deformation of adapters at 10 kHz. (a) Hexagonal adapter, (b) cylindrical adapter with cutting. Color maps show the deviation of displacement amplitude from the average value of six-measurement points for the hexagonal adapter.

For the hexagonal adapter, the displacement amplitude distributes concentrically (Fig. 6(a)). The deviation range is approximately −0.6% to +0.7%, and the position closer to the center, the larger the amplitude, implying that the accelerometer "pushes and pulls" the adapter with an inertial force. Consequently, the stress deforms the adapter.

Comparably, the result for the cylindrical adapter (Fig. 6(b)) obviously shows a distribution along with the cutting direction. It suggests that the inertial force bends the adapter along the cutting axis, unlike the hexagonal adapter. This asymmetry around the center causes systematic bias in the calibration results when measurement positions are inadequately controlled. The deviation ranges from −1.2% to +0.9%. Thus, calibration results may be affected at this level for the largest case. Note that the resonant frequencies of the adapters of similar shapes sufficiently exceed 10 kHz [13]. Thus, this deformation is not a resonant mode, but a quasi-static deformation.

Here, we roughly estimated the displacement deviation magnitude for the cylindrical adapter using the material's elasticity with a simplified bar model (Fig. 7). We considered a both-ends supported bar with an equally distributed load at the center. From a formula in material mechanics, the displacement amplitude $\delta$ induced by the accelerometer's inertia is written as

$$\delta = \frac{(8L^3 - 4Lb^2 + b^3)F}{384El}, \quad (1)$$

where $F$ is the inertial force and $E$ is Young's modulus (modulus of longitudinal elasticity). As shown in Fig. 7, $L$ and $b$ represent the length of the bar and the width of the applied force, respectively. The moment of inertia of area $l$ is calculated as $l = \frac{WH^3}{12}$, where $W$ and $H$ is the width and the height of the bar, respectively. The inertial force $F$ is written as $F = ma$, where $m$ is the accelerometer's mass, $a$ is the vibration amplitude in acceleration. Considering that the vibration amplitude in displacement $d$ satisfies $d = a \cdot (2\pi f_{\text{vib}})^{-2}$, we estimated the deviation ratio of the displacement $\delta/d$ as

$$\frac{\delta}{d} = \frac{(8L^3 - 4Lb^2 + b^3)}{32WH^3} \frac{m(2\pi f_{\text{vib}})^2}{E}. \quad (2)$$

Notably, equation (1) shows that the deviation ratio does not depend on the acceleration amplitude.

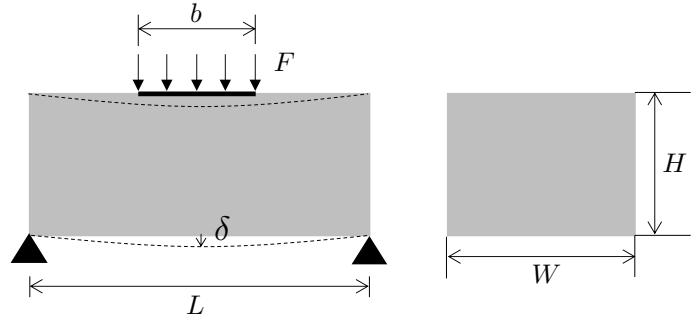

Fig. 7. Simplified model for rough estimation of deflection of the cylindrical adapter.

Substituting the following parameters as $m = 26$ g, $L = 22$ mm, $H = 12$ mm, $W = 19$ mm, $E = 193$ GPa, $b = 13.0$ mm, and $f_{\text{vib}} = 10$ kHz, we derived $\delta/d = 2.4\%$. The observed value (from $-1.2\%$ to $+0.9\%$) is roughly consistent with the estimated value. Note that the simplified model differs from the actual shape of the adapter. Thus, this estimation is only an order evaluation. For precise estimation, a numerical simulation, e.g., a finite element method (FEM) analysis, is needed.

Fig. 8 shows the radial dependence of the deviation of displacement amplitude. Each measurement point in Fig. 6(a) is replotted with the distance from the center. Here, the radius of the six-measurement points used for the origin (0% deviation) is 10.5 mm. Around that radius, the derivative is approximately 0.2%/mm. This position-dependence factor is useful to estimate the uncertainty components regarding positioning uncertainty; 1-mm uncertainty in manual alignment results in 0.2% uncertainty component at 10 kHz, which is not negligible. Notably, larger uncertainty may arise when the cylindrical adapter is used.

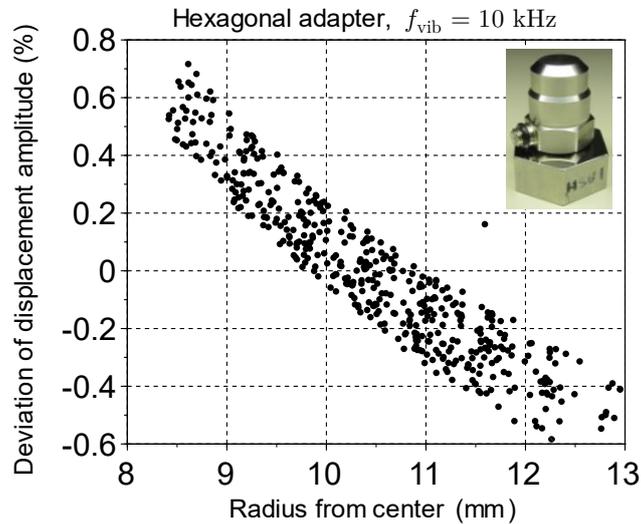

Fig. 8. Deviation of displacement amplitude as a function of radius from the center.

## 4. DISCUSSION

In this study, a two-axis precision linear stage was applied to move the heterodyne laser interferometer's head. The positioning stage offers better positioning repeatability of 1 μm, which is impossible with manual alignment mechanisms used in commercial primary accelerometer calibration systems. Compared with the galvanometer scanner, which has a 26-μm positioning resolution [3], better positioning capability is also achieved. Note that the repeatability of 1 μm is not the total positioning accuracy of the system, which may be deteriorated by other factors, such as instability between the positioning stage and the vibration exciter.

The position-dependent calibration results (Fig. 6) indicate that the precise, deterministic, and reproducible positioning of the laser beam is critical in evaluating the equivalency of different calibration systems. In previous international comparisons, the transfer artifacts were designated as the combined mechanical body of the accelerometer and adapter to avoid the influence of different vibration exciters [14,15]. However, Figs. 6 and 8 imply that the calibration results may still change when a different position on the adapter is measured; the measurement position should be strictly specified for consistent results. Besides, the measured sensitivity differs from the "true" sensitivity of the single-ended accelerometer because no one can access the mounting surface of the accelerometer directly with a laser beam. This issue is not yet solved by the proposed system.

The system enabled us to perform a vibration resonance analysis or deformation analysis efficiently, owing to the reduced lead time of calibrations; it requires 128 min per 466 measurement positions in this system. Analysis of deformations and resonances is useful for improving the accelerometer calibration, especially at high frequencies. With the conventional single-point calibration, conversely, the deformation can only be shown as a large discrepancy among measurement positions or poor repeatability. Note that the lead time can be improved further because more than half of the measurement time is consumed by waiting time between measurements and signal processing in the present configuration.

The visualization of surface deformation of the adapters also has implications on what is the ideal type of adapter. Obviously, a greater Young's modulus $E$ is beneficial. Tungsten carbide ($E \simeq 600$ GPa) can reach more than twice the Young's modulus of stainless steel, thus the magnitude of the deformation may be halved. The shape factor of the adapter also has a major impact on the deformation. For example, thicker adapter (greater $H$) may be beneficial, according to equation (2). Besides, a hexagon seems to be the best, having both appropriate symmetry and convenience of handling.

In terms of measurement error, the proposed system is free of additional cosine errors. This is because the laser beam angle is unchanged from the perpendicular direction, unlike galvanometer scanner. Additionally, this system can also operate with a collimated beam, instead of the focused beam presented herein. With the collimated beam, the measurement target can move in a wide range, up to several meters. Thus, the positioning stage can also be applied in a low-frequency accelerometer calibration system, which requires large displacement.

In the proposed system, the positioning stage was applied to the laser interferometer, not to the vibration exciter. In industrial measurement and inspection systems, positioning stages are normally installed to move DUTs, rather than measuring instruments. However, we chose this configuration not to excite the resonant mode of the vibration exciter. The heavyweight of the vibration exciter (>20 kg, including jigs) also inhibits the use of the precise positioning stages. In contrast, the laser interferometer is lightweight, and at the same time, robust to noise, as long as the stage is stopped. We also confirmed the long-term reliability of the laser interferometer. No technical issue was encountered on the laser interferometer for more than three months, with the measurements at several thousand positions in total.

The automatic positioning stage can be applied to other systems for vibration and shock metrology, especially when the reference surface of the calibration has nonuniform motion. For example, primary shock calibration suffers from surface deformation [16]. Thus, the positioning stages can be applied to our primary shock calibration system [17]. Nonrectilinear motions in low-frequency (<~300 Hz) primary vibration systems can also be distinguished and even compensated using the proposed system.

## 5. CONCLUSION

Herein, a primary accelerometer calibration system using a two-axis precise positioning stage to move the laser interferometer was developed. The proposed system offers an efficient primary calibration of accelerometers. The repeatability of single-point measurement was 0.001% – 0.1% in the frequency range of 5 Hz – 20 kHz. To take the average of six-point measurements, the repeatability was kept below 0.1%, even up to 20 kHz. Utilizing the precise positioning capability, surface deformations of different adapters at 10 kHz were visualized, implying that the deformation is caused by the accelerometer's inertial force. Moreover, we found that the hexagonal-shaped adapter is better than the cylindrical adapter with cutting. Precise control and reproducibility of the measurement position are essential for consistent comparison of single-ended accelerometers, such as the international comparison of national metrology institutes.

## ACKNOWLEDGMENTS

The authors acknowledge Hiromi Mitsumori, Yoshiteru Kusano, and Akihiro Ota (NMIJ/AIST) for their help and advice.

## REFERENCES

[1] ISO 16063-11:1999 Methods for the calibration of vibration and shock transducers—Part 11: Primary vibration calibration by laser interferometry (1999).

[2] T. Bruns, et al., "Final report on the CIPM key comparison CCAUV.V-K5," Metrologia **58**(1A), 09001 (2021).

[3] D. Sprecher, and H. Christian. "Primary accelerometer calibration by scanning laser Doppler vibrometry." Measurement Science and Technology **31**, 065006 (2020).

[4] A. Oota, et al. "Development of primary calibration system for high frequency range up to 10 kHz," Proc. Of IMEKO 20th TC3, 3rd TC16 and 1st TC22 International Conference (2007).

[5] W. Kokuyama, T. Shimoda, and H. Nozato, "An automated multipoint primary vibration calibration system," Measurement: Sensors **18**, 100140 (2021).

[6] W. Kokuyama, H. Nozato, and T.R. Schibli. "Phase meter based on zero-crossing counting of digitized signals," arXiv preprint arXiv:2009.01137 (2020).

[7] P.L. Heydemann, "Determination and correction of quadrature fringe measurement errors in interferometers," Applied Optics **20**, 3382-3384 (1981).

[8] W. Kokuyama, T. Shimoda, and H. Nozato, "Vibration measurement without the Heydemann correction," Measurement: Sensors **18**, 100136 (2021).

[9] JIS B0601 (Geometrical Product Specifications (GPS) -- Surface texture: Profile method -- Terms, definitions, and surface texture parameters.) Japan Industrial Standard, https://www.jisc.go.jp/

[10] T. Shimoda, W. Kokuyama, and H. Nozato, "Precise sinusoidal signal extraction from noisy waveform in vibration calibration," arXiv preprint, arXiv:2203.12144 (2022).

[11] ISO 266:1997 Acoustics—Preferred frequencies (1997).

[12] G. P. Ripper, R. S. Dias, and G. A. Garcia, "Primary accelerometer calibration problems due to vibration exciters," Measurement **42**, 1363-1369 (2009).

[13] G. P. Ripper, G. B. Micheli, and R. da Silva Dias, "A proposal to minimize the dispersion on primary calibration results of single-ended accelerometers at high frequencies," Acta IMEKO **2**, 48-55, (2014).

[14] A. Täubner, H. Schlaak, M. Brucke, and T. Bruns, "The influence of different vibration exciter systems on high frequency primary calibration of single-ended accelerometers," Metrologia **47**, 58, (2009).

[15] T. Bruns, A. Link, and A. Täubner, "The influence of different vibration exciter systems on high frequency primary calibration of single-ended accelerometers: II," Metrologia **49**, 27, (2011).

[16] H. Volkers, T. Beckmann, and R. Behrendt. "Investigations of reference surface warp at high shock calibrations." Proc. of Joint IMEKO International TC3, TC5 and TC22 Conference. (2014).

[17] H. Nozato, T. Usuda, A. Oota, and T. Ishigami, "Calibration of vibration pick-ups with laser interferometry, part IV. Development of a shock acceleration exciter and calibration system," Measurement Science and Technology **21**, 065107 (2010).